\documentclass[%
 aip,
 amsmath,amssymb,
 reprint,%
]{revtex4-1}

\usepackage[dvipdfmx]{graphicx}
\usepackage{dcolumn}% Align table columns on decimal point
\usepackage{bm}% bold math
\usepackage[utf8]{inputenc}
\usepackage[T1]{fontenc}
\usepackage{mathptmx}
\usepackage{etoolbox}

\usepackage{color} 
\usepackage{ulem} 
\usepackage{siunitx}

\makeatletter
\def\@email#1#2{%
 \endgroup
 \patchcmd{\titleblock@produce}
  {\frontmatter@RRAPformat}
  {\frontmatter@RRAPformat{\produce@RRAP{*#1\href{mailto:#2}{#2}}}\frontmatter@RRAPformat}
  {}{}
}%
\makeatother
\begin{document}

\preprint{AIP/123-QED}

\title{Surface barrier effect as evidence of chiral soliton lattice formation in chiral dichalcogenide CrTa$_{3}$S$_{6}$ crystals}
% Force line breaks with \\

\author{K. Mizutani}
\affiliation{Department of Physics and Electronics, Osaka Metropolitan University, Sakai, Osaka 599-8531 Japan}

\author{J. Jiang}
\affiliation{Department of Physics and Electronics, Osaka Metropolitan University, Sakai, Osaka 599-8531 Japan}

\author{K. Monden}
\affiliation{Department of Physics and Electronics, Osaka Metropolitan University, Sakai, Osaka 599-8531 Japan}

\author{Y. Shimamoto}
\affiliation{Department of Physics and Electronics, Osaka Metropolitan University, Sakai, Osaka 599-8531 Japan}

\author{Y. Kousaka*}
\email{koyu@omu.ac.jp}
\affiliation{Department of Physics and Electronics, Osaka Metropolitan University, Sakai, Osaka 599-8531 Japan}

\author{Y. Togawa}
\affiliation{Department of Physics and Electronics, Osaka Metropolitan University, Sakai, Osaka 599-8531 Japan}

\date{\today}% It is always \today, today,
             %  but any date may be explicitly specified

\begin{abstract}
The formation of chiral magnetic soliton lattice (CSL) is investigated in monoaxial chiral dichalcogenide CrTa$_{3}$S$_{6}$ crystals in terms of a surface barrier, which prevents a penetration of chiral solitons into the system and is an intrinsic origin of hysteresis for the continuous phase transition of nucleation-type, as discussed in the system of quantized vortices in type-II superconductors. The magnetoresistance (MR) was examined with microfabricated platelet samples in different dimensions with regard to the $c$-axis direction of the crystal. The CSL formation was confirmed by the discrete MR changes, reflecting the number of chiral solitons, as well as by the presence of surface barrier, recognized as a fixed ratio of critical magnetic fields during the hysteresis field cycle. We also argue the influence of the surface barrier in the bulk CrTa$_{3}$S$_{6}$ crystals.
\end{abstract}

\maketitle

\section{Introduction}

Chiral helimagnets induce an antisymmetric exchange interaction strongly coupled to a chiral crystalline structure~\cite{Dzyaloshinskii1958, Moriya1960}. As a consequence of its competition with a symmetric Heisenberg exchange interaction in the absence or presence of magnetic fields, nontrivial chiral magnetic structures emerge such as chiral helimagnetic order (CHM)~\cite{Dzyaloshinskii1958, Moriya1960}, chiral soliton lattice (CSL)~\cite{Dzyaloshinskii1964, Dzyaloshinskii1965a, Dzyaloshinskii1965b, Izyumov1984, Kishine2005}, and chiral magnetic vortices called magnetic Skyrmions~\cite{Bogdanov1989, Bogdanov1994}. 
These chiral magnetic structures have been observed via neutron scattering or electron microscopy in a recent decade~\cite{Muhlbauer2009, Yu2010, Togawa2012}.

\begin{figure}[tb]
\begin{center}
\includegraphics[width=8.5cm]{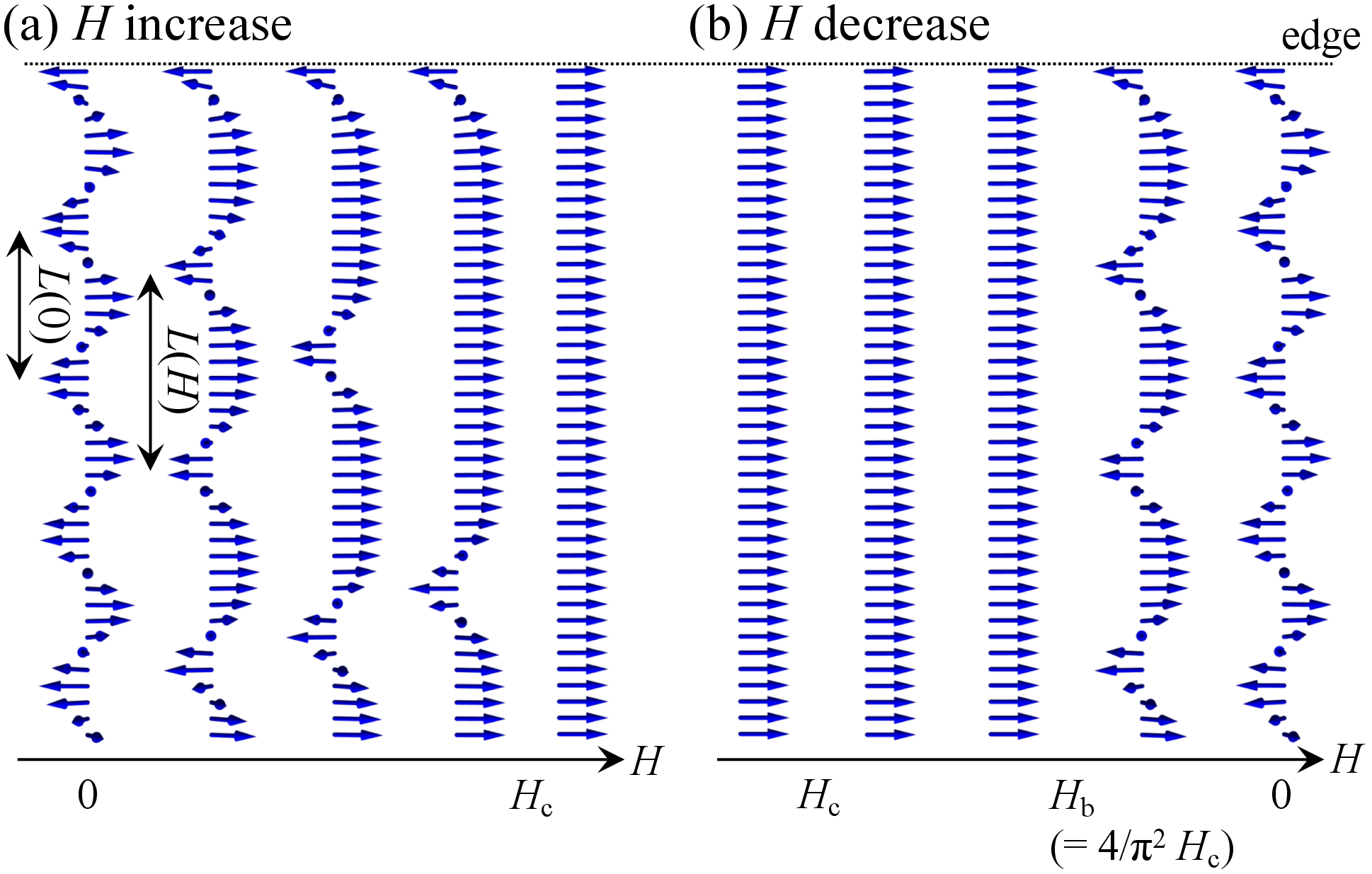}
\end{center}
\caption{
The formation of chiral soliton lattice (CSL) in the $H$ increase (a) and decrease (b) processes in a semi-infinite system with a boundary between the material and vacuum. The CSL undergoes a continuous phase transition to a forced ferromagnetic state toward $H_{\rm c}$ in the former case, while the surface barrier prevents the penetration of chiral solitons until it disappears at $H_{\rm b}$ in the latter case. There is a chiral surface twist structure at the sample edge, the formation of which is associated with the presence of the surface barrier.
}
\label{f-str}
\end{figure}

CrNb$_3$S$_6$ is one of the well-established transition-metal dichalcogenides (TMD) that exhibit chiral helimagnetism. 
CrNb$_3$S$_6$ forms a chiral monoaxial crystal structure with space group $P6_{3}22$~\cite{Berg1968, Anzenhofer1970, Laar1971}, where the symmetric and antisymmetric exchange interactions work along the principal $c$-axis of the crystal.
The CSL formation, as schematically drawn in Fig.~\ref{f-str}(a), was detected in real space images and reciprocal scattering data by using Lorentz microscopy in CrNb$_3$S$_6$~\cite{Togawa2012}. 
Neutron~\cite{Miyadai1983} and resonant magnetic X-ray~\cite{Honda2020} scattering experiments are also useful for identifying the CHM and CSL in a reciprocal space.

The CSL exhibits robust phase coherence at macroscopic length scale~\cite{Togawa2012}.
Thus, nontrivial characteristics appear in various physical properties.
Indeed, the CSL shows giant magnetoresistance (MR) due to a proliferation of chiral solitons~\cite{Togawa2013}, discretization MR effect~\cite{Togawa2015}, robust response of chiral solitons under oblique magnetic fields~\cite{Yonemura2017}, nonreciprocal electrical transport~\cite{Aoki2019}, and collective elementary excitation of the CSL up to a frequency of sub-terahertz~\cite{Shimamoto2022}.
Moreover, very anisotropic soliton defects appear in the CSL system when decreasing the magnetic field $H$~\cite{Gary2019,Goncalves2020,Li2023}.
Such coherent, topological, and collective nature of the CSL could be useful for spintronic device applications such as memory multi-bits and 6G communications technology using the CSL~\cite{Kishine2015, Togawa2016, Togawa2023}.

The MR is one of the feasible methods for identifying the CSL. In particular, when reducing the sample dimensions, discretized MR appears because of a countable nature of chiral solitons in the CSL~\cite{Togawa2015}.
In addition,
the surface barrier emerges upon the penetration of chiral solitons into the system in the $H$ decrease process, as schematically drawn in Fig.~\ref{f-str},
because of the phase coherence of the CSL~\cite{Shinozaki2018}.
The strength of surface barrier is quantified by solving a one-dimensional (1D) chiral sine-Gordon model in a semi-infinite system, which describes well the CSL system realized in the monoaxial chiral helimagnets such as CrNb$_3$S$_6$. 
The ratio of the magnetic field $H_{\rm b}$,
where the surface barrier disappears, to the critical magnetic field $H_{\rm c}$ takes a constant value ($H_{\rm b}$/$H_{\rm c}$ = $4/\pi^2 \sim 0.405)$~\cite{Shinozaki2018}.
In the experiments, the presence of surface barrier could be detected as a sudden jump of physical quantity at a particular strength of the magnetic field ($H_{\rm jump}$) when decreasing $H$. The values of $H_{\rm jump}$/$H_{\rm c}$, which were experimentally obtained by using the MR in CrNb$_3$S$_6$ (e.g., 0.416, 0.405, and 0.408 in a particular micrometer-sized crystal)~\cite{Shinozaki2018}, showed an excellent agreement with $4/\pi^2$ expected for the 1D chiral sine-Gordon model. 
This coincidence is regarded as evidence of the presence of CSL. 
Importantly, the surface barrier is an intrinsic origin of hysteresis for the CSL system that exhibits continuous phase transition of nucleation-type~\cite{deGennes}, as discussed in the Bean-Linvingston barrier~\cite{BL1964, deGennes_textbook} for Abrikosov quantized vortices in type-II superconductors.
The surface barrier has been evaluated quantitatively for the first time in the CSL system in CrNb$_3$S$_6$~\cite{Shinozaki2018}. The importance of the surface barrier and related surface twist structure was also discussed in cubic chiral helimagnets in the study of magnetic Skyrmions~\cite{Wilson2013, Meynell2014}.

Recently, CrTa$_3$S$_6$ and related TMDs, which form the same crystal structure as that of CrNb$_3$S$_6$, have also attracted attention because of the possible emergence of chiral helimagnetism~\cite{Kousaka2016, Zhang2021, Obeysekera2021}. 
CrTa$_3$S$_6$ was initially reported as a ferromagnetic compound~\cite{Parkin1980a, Parkin1980b}.
However, reexamination has revealed that CrTa$_3$S$_6$ exhibits the CHM and CSL~\cite{Kousaka2016, Zhang2021, Obeysekera2021}.
Now, it turns out that CrTa$_3$S$_6$ has a helimagnetic period of \SI{22}{\nm} and large $H_{\rm c}$ of \SI{1.2}-\SI{1.6}{\tesla}, while \SI{48}{\nm} and \SI{0.2}{\tesla} respectively in CrNb$_3$S$_6$. The discretization MR effect was observed in the cleaved CrTa$_3$S$_6$ samples~\cite{Zhang2022}, as reported in CrNb$_3$S$_6$~\cite{Wang2017}. However, there has been no experimental report on the surface barrier effect in CrTa$_3$S$_6$. It is not clear whether the surface barrier works among the chiral helimagnets hosting the CSL.

In this paper, we investigate magnetic and transport properties of CrTa$_3$S$_6$ in the viewpoint of the surface barrier effect during the CSL formation. To scrutinize this unique property, the MR and magnetization measurements were performed with micrometer-sized and bulk crystals with different dimensions with regard to the $c$-axis direction. The obtained results demonstrate the existence of surface barrier in CrTa$_3$S$_6$ crystals. Namely, characterizing the surface barrier is useful for identifying the CSL system in chiral magnetic materials.

\section{Experimental Methods}

Single crystals of CrTa$_{3}$S$_{6}$ were obtained by chemical vapor transport (CVT) technique in a temperature gradient using iodine I$_{2}$ as a transporting agent \cite{Miyadai1983, Kousaka2009}.
The polycrystalline powders, synthesized by gas phase method with a mixture of Cr, Ta and S in the molar ratio of $x_{\rm nominal} : 3 : 6$
($x_{\rm nominal}$ is the nominal amount of Cr),
were placed at one end of an evacuated silica tube and then heated in the electric tube furnace under the fixed temperature gradient from \SI{1100}{\degreeCelsius} to \SI{1000}{\degreeCelsius} for two weeks.
The bulk crystals were grown at the other end of the silica tube. The grown crystals have the shape of a hexagonal plate of around \SI{0.5} to \SI{1.0}{\mm} in diameter and of \SI{100}{\um} in thickness.

The magnetization of the obtained bulk crystals was examined using a SQUID magnetometer (Quantum Design MPMS3).
Magnetoresistance (MR) measurements were performed with the micrometer-sized specimens of CrTa$_3$S$_6$ crystals, which were prepared from the bulk CrTa$_3$S$_6$ crystal by using a focused ion beam (FIB) system~\cite{Aoki2019}.
The size of the specimens was evaluated using a scanning electron microscopy system.
The MR data were collected by the standard four-terminal method using a physical property measurement system (Quantum Design PPMS).
Note that $H$ was applied in the direction parallel to the sample plane of bulk and microfabricated crystals, as described below, so as to reduce demagnetizing field and extrinsic metastability effects~\cite{Mito2018}.

Three configurations of the platelet specimens with regard to the $c$-axis direction were prepared for the MR measurements.
In the first case, a $c$-plane sample was fabricated, as shown in Fig.~\ref{f-MR1}(a). A dimension of this platelet sample \#1 is \SI{12}{\um} $\times$ \SI{4}{\um} $\times$ \SI{0.1}{\um}, where the shortest length corresponds to the direction of the $c$-axis. This length limits the maximum number of the solitons in the CSL and thus the discretization effect was observed in the present CrTa$_3$S$_6$ crystal, as seen in CrNb$_3$S$_6$~\cite{Togawa2023}.
Another platelet samples \#2 and \#3 were fabricated with the $c$-axis being along the longitudinal direction of the sample plane, as shown in Figs.~\ref{f-MR1}(b) and~\ref{f-MR1}(c).
A size of the sample \#2 is \SI{9}{\um} $\times$ \SI{1}{\um} $\times$ \SI{10}{\um}, where the longest length corresponds to the $c$-axis direction. This sample shape is similar to that of CrNb$_3$S$_6$ used for the demonstration of the surface barrier~\cite{Shinozaki2018}.
The sample \#3 has an elongated geometry with a dimension of \SI{2}{\um} $\times$ \SI{1}{\um} $\times$ \SI{19}{\um} (// $c$-axis).

\begin{figure}[tb]
\begin{center}
\includegraphics[width=8.5cm]{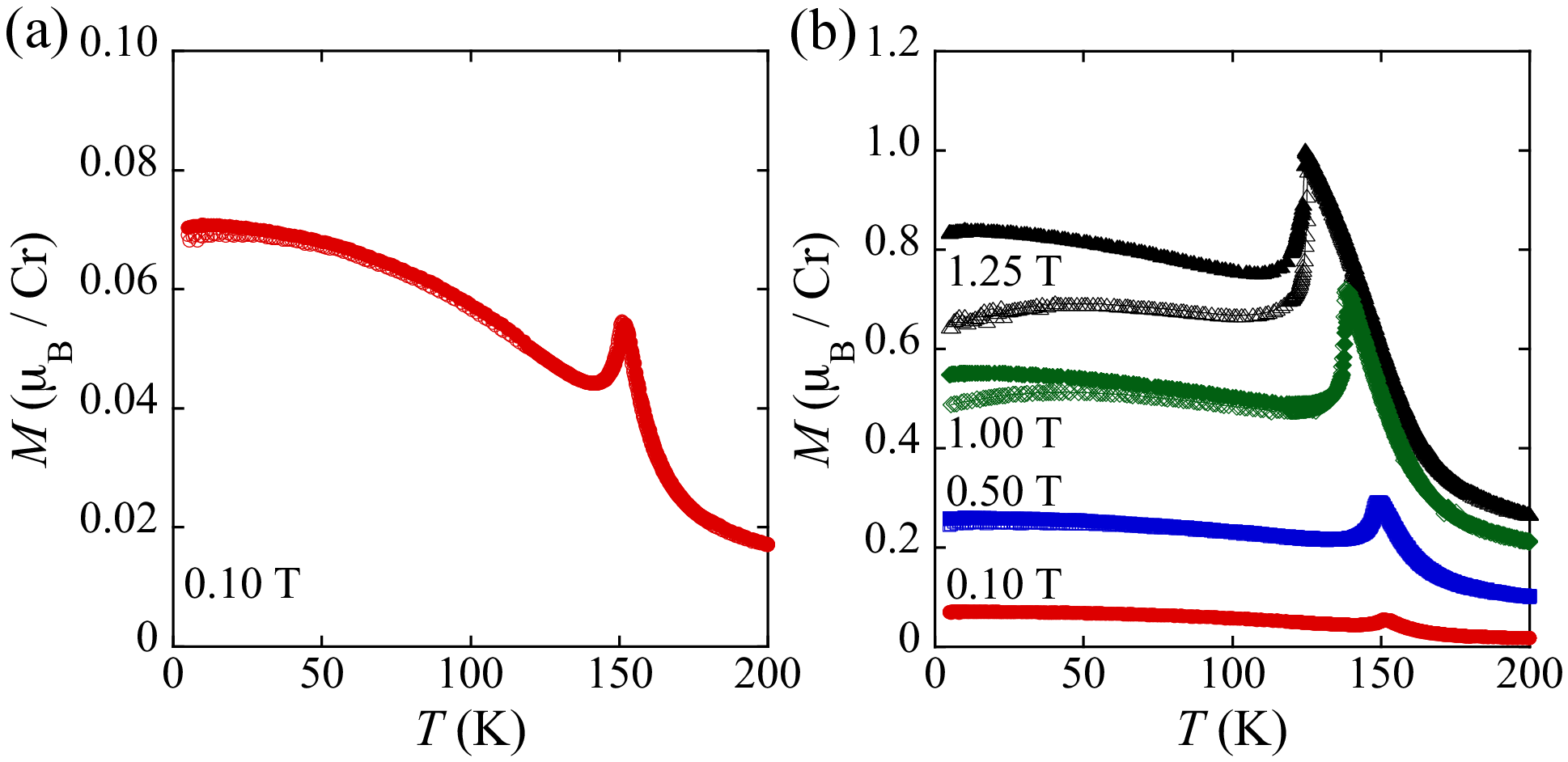}
\end{center}
\caption{
Temperature dependence of the magnetization in the CrTa$_{3}$S$_{6}$ single crystal at \SI{0.10}{\tesla} (a) and at higher $H$s up to \SI{1.25}{\tesla} (b). $H$ was applied in the direction perpendicular to the $c$-axis. Closed and open marks denote the magnetization data collected in the field cooling and zero-field cooling processes, respectively.
}
\label{f-MT}
\end{figure}

\section{Experimental Results}

\begin{figure*}[tb]
\begin{center}
\includegraphics[width=12.8cm]{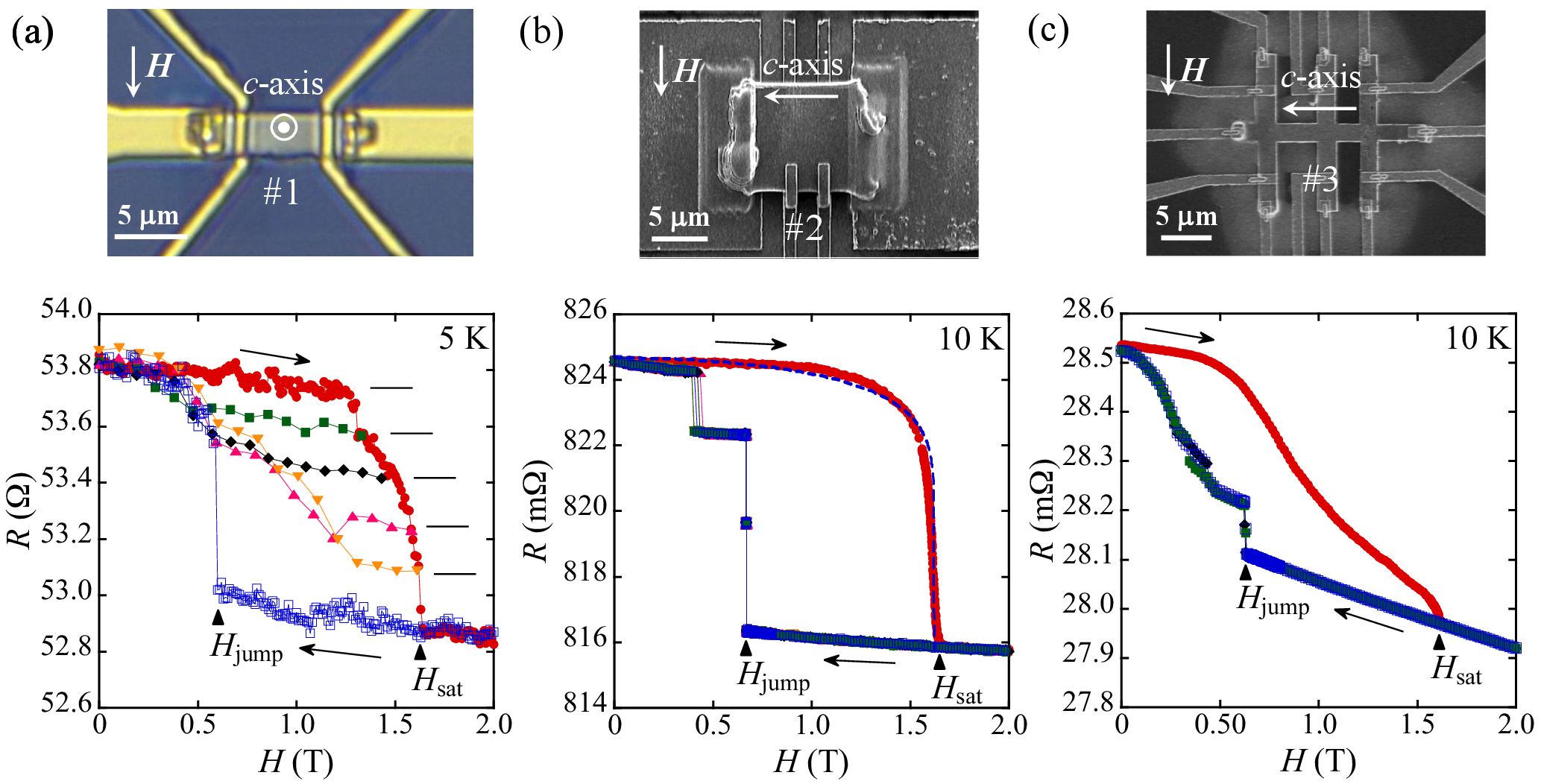}
\end{center}
\caption{
MR data taken in three micrometer-sized platelet CrTa$_{3}$S$_{6}$ crystals. (a) Full and minor loops of the MR in the $c$-plane sample \#1 with $H$ applied in the direction perpendicular to the $c$-axis. The red closed circles represent the MR data during the $H$ increase process toward above $H_{\rm sat}$, while the other symbols correspond to the MR data in the $H$ decrease process. The discretization effect  of the MR behavior is clearly observed. 
(b and c) MR data in the samples with the $c$-axis pointing out along the longitudinal direction of the platelet sample.
The $c$-axis lengths of the samples \#2 (b) and \#3 (c) are \SI{10}{\um} and \SI{19}{\um}, respectively.
The red closed circles represent the MR data in the $H$ increase process, and the blue dotted line in (b) corresponds to a theoretical equation of the soliton density. The other symbols show the MR data in the $H$ decrease process taken five and three times repeatedly for the samples \#2 and \#3, respectively.
The ratio $H_{\rm sat}/H_{\rm jump}$ turns to be 0.407 for the sample \#2 and 0.392 for \#3 on average.
These results indicate that the surface barrier works against the penetration of chiral solitons.
}
\label{f-MR1}
\end{figure*}

For determining the optimum condition for the crystal growth, it should be noted that the magnetic property of the grown crystals is very sensitive to $x_{\rm nominal}$ of the powder precursor.
For instance, in the case of CrNb$_3$S$_6$ crystal growth~\cite{Kousaka2022}, the amount of Cr directly measured in the single crystals was found to be smaller than $x_{\rm nominal}$.
The $x_{\rm nominal}$ was determined to be 1.11 so as to obtain the ideal crystals of CrNb$_3$S$_6$ without Cr defects.

The optimization of CrTa$_3$S$_6$ crystal growth was performed by using powder precursors with $x_{\rm nominal}$ from 1.00 to 1.50.  
Imprints of the CSL formation were successfully obtained in the crystals grown with $x_{\rm nominal}$ = 1.29, while ferromagnetic response appeared in other crystals with different $x_{\rm nominal}$ values.

Figure~\ref{f-MT}(a) shows a peak anomaly of the magnetization of the obtained CrTa$_3$S$_6$ crystal at around \SI{150}{\kelvin} with a magnetic field $H$ of \SI{0.1}{\tesla} applied in a direction perpendicular to the $c$-axis.
Here, the critical temperature of the helimagnetic phase transition $T_{\rm c}$ is defined at a peak top of the magnetization.
The $T_{\rm c}$ values decrease with increasing the $H$ strength, as shown in Fig.~\ref{f-MT}(b).
Note that the $T_{\rm c}$ of \SI{150}{\kelvin} is \SI{10}{\kelvin} higher than those reported in the previous studies \cite{Zhang2021, Obeysekera2021}.
Such a variation of the $T_{\rm c}$ values indicates that the crystals used in the present study may have a small amount of Cr defects, as reminiscent of a dome-shaped profile of the relationship between $T_{\rm c}$ and $x_{\rm nominal}$ discussed in CrNb$_3$S$_6$~\cite{Kousaka2022}.

To see the presence of chiral solitons in the obtained CrTa$_3$S$_6$ crystals via the discretization effect, the MR was examined in the $c$-plane thin sample with $H$ applied in the direction perpendicular to the $c$-axis. 
First, the MR full loop was taken by cycling $H$ between zero and above the critical magnetic field (defined as $H_{\rm sat}$ in the experiments),
where all the chiral solitons escape from the sample and magnetic moments are likely to be saturated. Then, the MR minor loops were collected by sweeping $H$ below $H_{\rm sat}$.

All the MR data are presented in the same panel in Fig.~\ref{f-MR1}(a). It is clear that, in the $H$ increase process of
the MR full loop, the MR exhibits a gradual negative change associated with a reduction of the number of chiral solitons,
whereas it shows a sudden jump at $H_{\rm jump}$ in the $H$ decrease process. 
In addition, six discrete MR values appear in a series of the MR minor loops. 
Taking into consideration the helical period of \SI{22}{\nano \metre} in CrTa$_3$S$_6$~\cite{Kousaka2016}, the thickness of the present MR sample was calculated to be approximately \SI{110}{\nano \metre}, which is consistent with the value estimated from the device fabrication.

Another feature is that the ratio of $H_{\rm jump}/H_{\rm sat}$ is found to be 0.362. Although this value is slightly smaller than the theoretical value~\cite{Shinozaki2018} $4/\pi^2$, such a large hysteresis may indicate the influence of surface barrier against the penetration of chiral solitons into the present CrTa$_3$S$_6$ crystal.

The presence of surface barrier was first demonstrated in the micrometer-sized platelet CrNb$_3$S$_6$ crystals with the $c$-axis orienting within the plane, in which the experimental data of $H_{\rm jump}/H_{\rm sat}$ is in an excellent agreement with the theoretical value $4/\pi^2$.
The slight discrepancy found in Fig.~\ref{f-MR1}(a) may be ascribed to the difference in the sample geometry.
In this respect, it is worth examining the MR behavior in terms of the surface barrier in the CrTa$_3$S$_6$ sample with the dimensions similar to those of the CrNb$_3$S$_6$ sample used in the previous study~\cite{Shinozaki2018}.

Figure~\ref{f-MR1}(b) shows that such a CrTa$_3$S$_6$ sample (\#2) indeed shows the MR hysteresis behavior. Note that $H$ is applied in the direction perpendicular to the $c$-axis and within the plane so as to eliminate the demagnetization effect. 
To precisely determine the $H_{\rm jump}$ and $H_{\rm sat}$ values, the MR measurements were performed five times repeatedly.
 
In the $H$ increase process, the MR change is well fitted by the CSL density, which is derived from the chiral sine-Gordon model and plays a role as an order parameter of the CSL formation. On the other hand, in the $H$ decrease process, the MR shows a sudden change at $H_{\rm jump}$.
Note that the position of $H_{\rm jump}$ and the amplitude of the MR change at $H_{\rm jump}$ were reproducible in all the five MR measurements. 
Moreover, the ratio $H_{\rm jump}/H_{\rm sat}$ was averaged to be 0.407, which is quite consistent with the theoretical value $(4/\pi^2 \sim 0.405)$. 
These features are consistent with those observed in the CrNb$_3$S$_6$ sample~\cite{Togawa2015, Togawa2016, Shinozaki2018}.

Figure~\ref{f-MR1}(c) shows the MR data in the sample \#3 with the elongated geometry. The data was collected three times.
The sudden change of the MR occurs at $H_{\rm jump}$ reproducibly.
The ratio $H_{\rm jump}/H_{\rm sat}$ was averaged to be 0.392, which is slightly smaller than the theoretical value.

The present MR data in the three different CrTa$_3$S$_6$ samples strongly supports that the surface barrier works against the penetration of chiral solitons. Namely, the CSL formation was successfully demonstrated in the CrTa$_3$S$_6$ crystals via the MR measurements.

\begin{figure}[tb]
\begin{center}
\includegraphics[width=8.5cm]{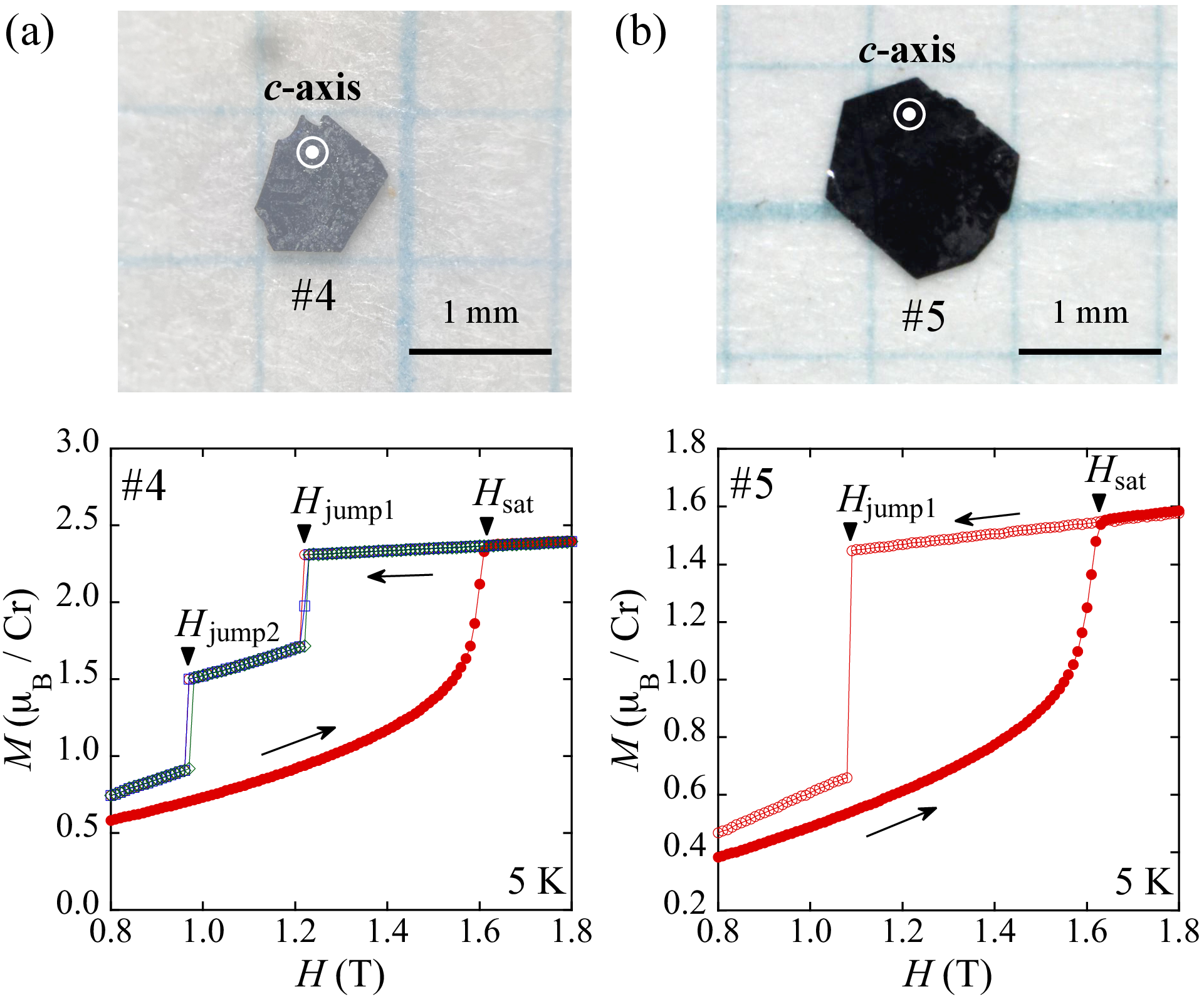}
\end{center}
\caption{
Magnetization curves at \SI{5}{\K} with two different bulk CrTa$_3$S$_6$ crystals \#4 (a) and \#5 (b).
Closed and open marks denote the magnetization data collected in the $H$ increase and decrease processes, respectively.
Sharp drops of the magnetization appear below the saturation field $H_{\rm sat}$ in the $H$ decrease process.
The first and second (last) jumps occur at almost the same $H$ value, which are respectively labeled as $H_{\rm jump1}$ and $H_{\rm jump2}$, in the crystal \#4, while a single large jump appears in the crystal \#5.
}
\label{f-MH1}
\end{figure}

\begin{figure}[tb]
\begin{center}
\includegraphics[width=8.5cm]{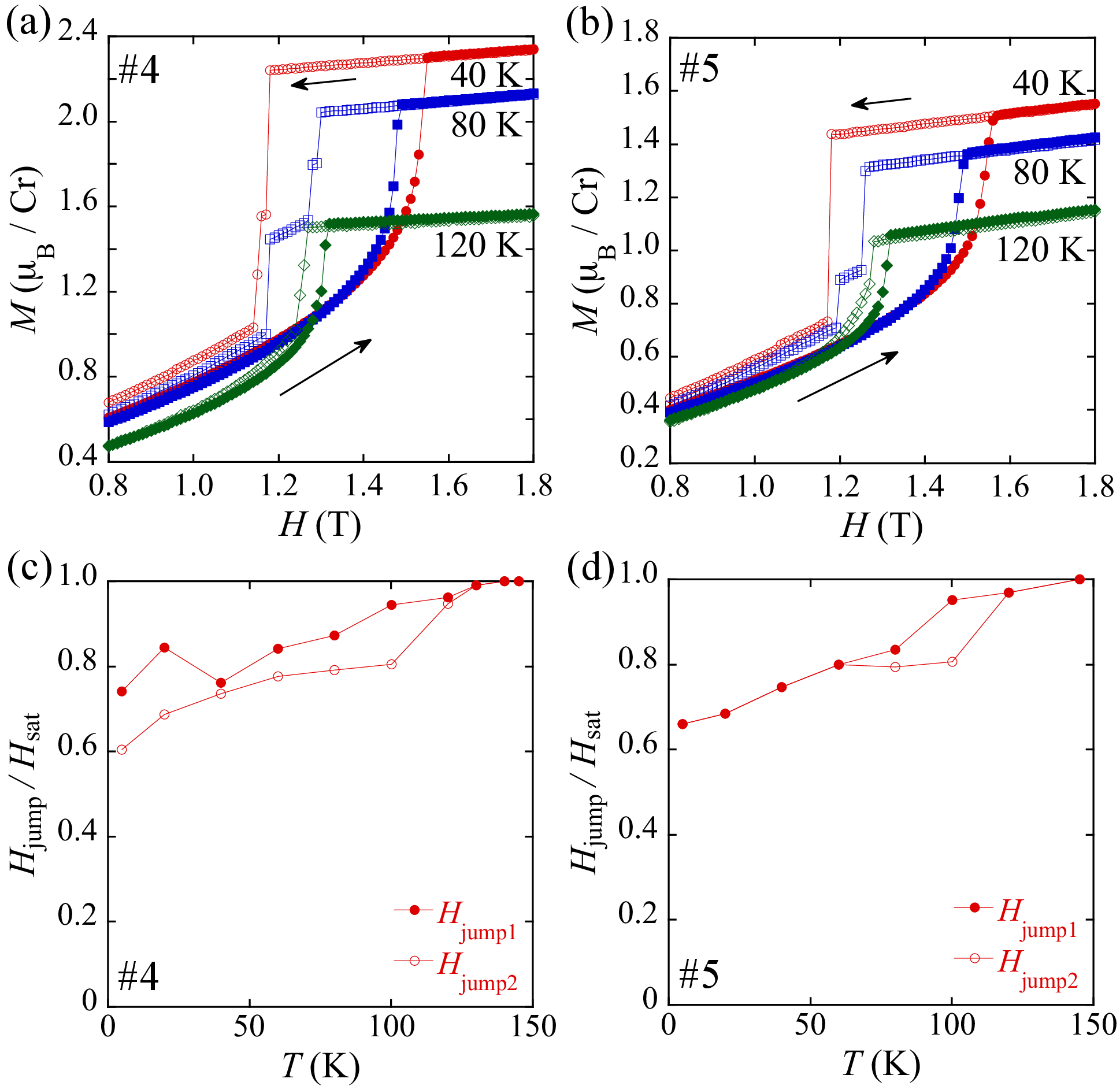}
\end{center}
\caption{
Magnetization curves at various temperatures with two different bulk CrTa$_3$S$_6$ crystals \#4 (a) and \#5 (b).
With increasing temperature, large hysteresis gradually shrinks in both crystals.
The $H_{\rm jump} / H_{\rm sat}$ values are given as a function of temperature for the crystals \#4 (c) and \#5 (d).
Here, $H_{\rm jump2}$ is identified as the last jump in the magnetization curves.
}
\label{f-MH2}
\end{figure}

\begin{figure}[tb]
\begin{center}
\includegraphics[width=8.5cm]{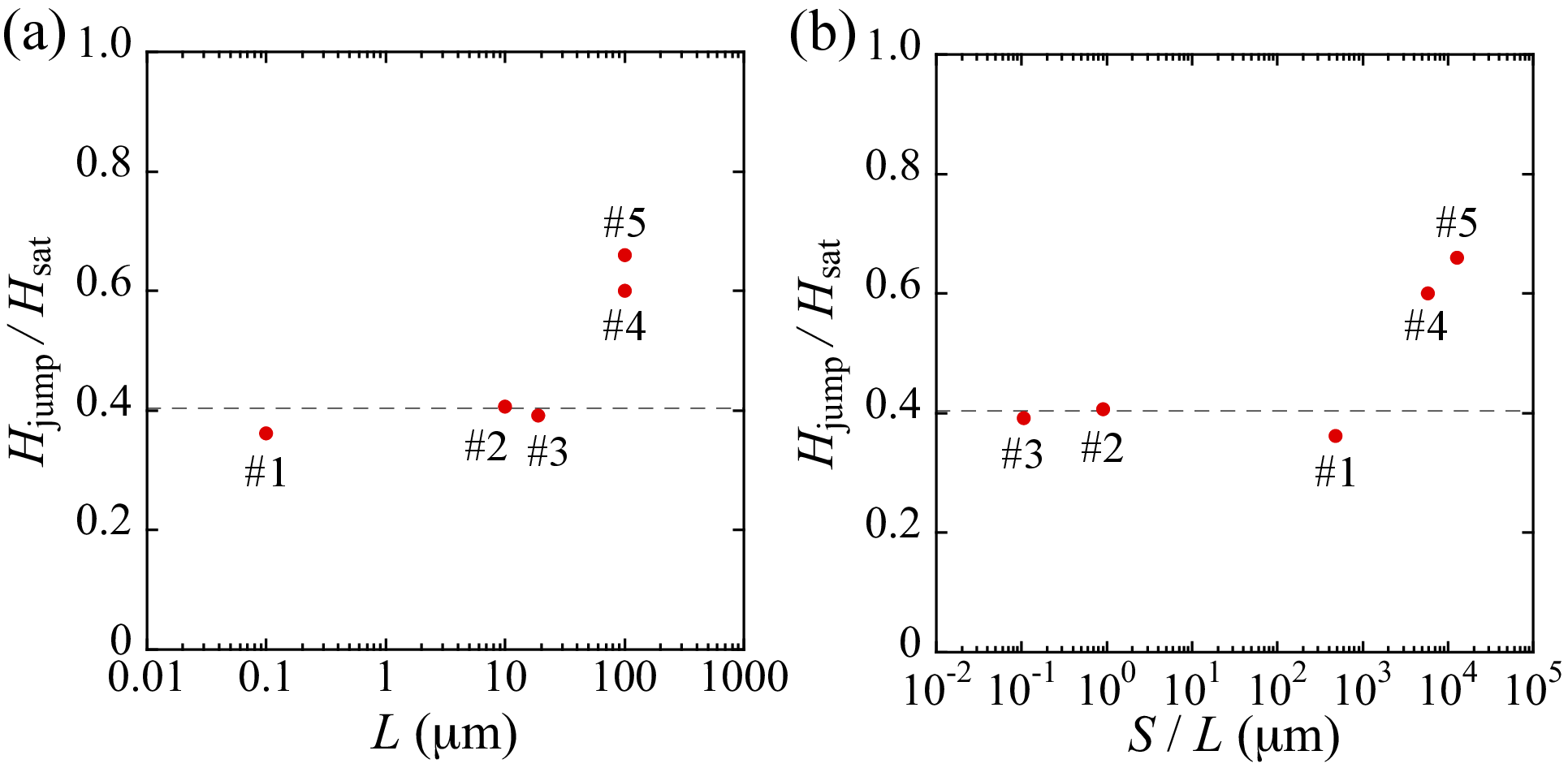}
\end{center}
\caption{
The ratio $H_{\rm sat}/H_{\rm jump}$ as a function of the sample geometry.
$H_{\rm sat}/H_{\rm jump}$ is evaluated in terms of the length $L$ along the $c$-axis in (a), while it is given with the $c$-plane area $S$ normalized by $L$ in (b). The numbers \#1 to \#5 correspond to the microfabricated and bulk samples shown in Figs.~\ref{f-MR1},~\ref{f-MH1} and~\ref{f-MH2}.
The dashed line represents the theoretical value $4/\pi^2$ of the surface barrier effect.
}
\label{f-surfacebarrier}
\end{figure}

Interestingly, the hysteresis behavior is observed in the magnetization curves even in the bulk CrTa$_3$S$_6$ crystals, as shown in Fig.~\ref{f-MH1}.
A typical geometry of the bulk crystals is a platelet shape with the $c$-plane being of about \SI{100}{\um} in thickness, as presented in the optical photographs in Fig.~\ref{f-MH1}. The magnetization curves at \SI{5}{\K} show the downward convex behavior in the $H$ increase process, which is regarded as evidence of the CSL formation~\cite{Dzyaloshinskii1965b, Izyumov1984, Kishine2005, Togawa2013, Togawa2016}.
On the other hand, in the $H$ decrease process, the magnetization decreases linearly until $H$ reaches down to the first $H_{\rm jump}$.
Sharp drops of the magnetization appear at $H_{\rm jump1}$ and $H_{\rm jump2}$ in the crystal \#4, while a drastic drop occurs at $H_{\rm jump1}$ in the crystal \#5.
The positions of $H_{\rm jump1}$ and $H_{\rm jump2}$ were confirmed to be reproducible in the repeated measurements.
This behavior is reminiscent of the MR behavior, as discussed in the micrometer-sized CrTa$_3$S$_6$ crystals in Fig.~\ref{f-MR1}, and indicates that the surface barrier hampers the penetration of chiral solitons into the bulk crystal.

To see an indication of the surface barrier, the dependence of the magnetization curves on temperature was examined, as shown in Figs.~\ref{f-MH2}(a) and \ref{f-MH2}(b).
The hysteresis remains small at temperatures in the vicinity of $T_{\rm c}$, while large hysteresis accompanying a sharp drop of the magnetization becomes evident with cooling temperature.

Figures~\ref{f-MH2}(c) and \ref{f-MH2}(d) shows the ratio $H_{\rm jump}/H_{\rm sat}$ as a function of temperature in the bulk CrTa$_3$S$_6$ crystals \#4 and \#5, respectively.
It is found that the $H_{\rm jump}/H_{\rm sat}$ values at \SI{5}{\K} reduce to $0.60$ and $0.66$ in the crystals \#4 and \#5, respectively.
These values are still larger than the theoretical value $4/\pi^2$ expected for the analytical model of surface barrier~\cite{Shinozaki2018}. 
Nevertheless, the behavior observed in the present CrTa$_3$S$_6$ crystals is totally different from
the magnetization data previously reported in bulk crystals of CrNb$_3$S$_6$~\cite{Tsuruta2015,Tsuruta2016} and CrTa$_3$S$_6$~\cite{Zhang2021}, where $H_{\rm jump}/H_{\rm sat}$ was respectively kept to be 0.82 -- 0.91 and 0.93 even at low temperatures.

The ratio $H_{\rm jump}/H_{\rm sat}$ is summarized in terms of the sample dimensions. 
For the crystals with the $c$-axis length of around ten micrometers, which contain hundreds of chiral solitons, $H_{\rm jump}/H_{\rm sat}$ shows a good agreement with the theoretical value, as seen in Fig.~\ref{f-surfacebarrier}(a).
When the $c$-plane area is normalized by the $c$-axis length, the experimental values tend to deviate from the theoretical value in the samples with large normalized area, as shown in Fig.~\ref{f-surfacebarrier}(b). In this respect, an elongated geometry along the helical axis is likely to be favorable for the surface barrier.

The surface barrier is an intrinsic effect in the CSL system with clean surface~\cite{Shinozaki2018}. 
In this respect, the discrepancy from the theoretical value may be ascribed to imperfect edges of the hexagonal-shaped bulk crystals, as shown in Fig.~\ref{f-MH1}, which is quite different from the ideal surface theoretically treated in the 1D chiral sine-Gordon model. 
Conversely, reproducible large hysteresis appears in the present bulk CrTa$_3$S$_6$ crystals. Namely, the chiral solitons in CrTa$_3$S$_6$ exhibit less extrinsic metastability than those in CrNb$_3$S$_6$, indicating that CrTa$_3$S$_6$ is an ideal material hosing the robust CSL.

The quality of the crystal may influence the effectiveness of surface barrier. 
Note that the $T_{\rm c}$ and $H_{\rm sat}$ values in the present CrTa$_3$S$_6$ crystals are larger than those reported in the literature. 
Indeed, $T_{\rm c}$ is \SI{10}{\K} higher than the reported values, as described above, and $H_{\rm sat}$ is \SI{1.65}{\tesla} at \SI{5}{\K}, which is \SI{0.35}{\tesla} larger than that in the previous work~\cite{Zhang2021}.
It was already found in CrNb$_3$S$_6$ that $T_{\rm c}$ and $H_{\rm sat}$ decrease when the amount of Cr intercalation deviates from the ideal unity~\cite{Kousaka2022} and are closely correlated with the strength of symmetric and antisymmetric exchange interactions. 
Importantly, the symmetric exchange interaction perpendicular to the $c$-axis $J_{\rm \perp}$ is enlarged in the present CrTa$_3$S$_6$ crystals because $T_{\rm c}$ is correlated with the strength of $J_{\rm \perp}$~\cite{Shinozaki2016}. $J_{\rm \perp}$ also works for enhancing the phase coherence of the CSL~\cite{Togawa2023}, which is in favor of working the surface barrier even in the bulk crystals.  
Clarifying the relationship between the strength of the exchange interactions and surface barrier would be an interesting issue in the CSL system.

The surface quality of the $c$-plane may also be a key element governing the strength of the surface barrier. In the experiments, the samples \#1 to \#3 were prepared by using FIB fabrication and the samples \#4 and \#5 have an as-grown wide $c$-plane surface. A clear difference was not found in terms of the surface quality but rather the controllability of the surface barrier was evident in the dependence on thickness and aspect ratio, as seen in Fig.~\ref{f-surfacebarrier}. The comparison of the surface barrier strength using various types of the surface such as a freshly-cleaved surface~\cite{Zhang2022, Wang2017} and a sharp crystal edge would promote the understanding of the surface barrier in the CSL system.

One of the unknown characteristics in the present CrTa$_3$S$_6$ crystals is a reduction of the magnetic moment at $H_{\rm sat}$ to 1.6 $\mu_{\rm B}/{\rm Cr}$, which is almost a half smaller than the value expected for the isolated Cr$^{3+}$ ion. The magnetic moment increases monotonically above $H_{\rm sat}$ and reaches to 3.0 $\mu_{\rm B}$ around \SI{10}{\tesla} with a linear extrapolation. This behavior is different from that reported in the previous work~\cite{Zhang2021}. The reason of such a discrepancy in CrTa$_3$S$_6$ remains to be clarified. 
Inferred from an electronic structure of CrNb$_3$S$_6$, itinerant electrons composed of Ta and S atoms may interact with localized electrons of Cr atoms in an electronic structure of CrTa$_3$S$_6$. 
However, the picture of localized electrons in the Cr atoms has not been fully validated yet. In this connection, it would be interesting to examine an electronic structure of CrTa$_3$S$_6$ using x-ray magnetic circular dichroism (XMCD) spectroscopy together with density functional theory (DFT) calculations to evaluate the degree of hybridization between Ta $5d$ and Cr $3d$ orbitals, as discussed in CrNb$_3$S$_6$~\cite{Mito2019}. 

In summary, we demonstrate the CSL formation via characterizing the surface barrier in CrTa$_3$S$_6$ crystals. This observation indicates that the surface barrier effect occurs among TMDs hosting the CSL and induces nontrivial physical response such as discretized MR reflecting the topological nature of the CSL.

\begin{acknowledgments}
We thank Yusuke Kato for fruitful discussion. This work was supported by JSPS KAKENHI Grant Numbers 19H05822, 19H05826, 23H01870 and 23H00091.
\end{acknowledgments}

\section*{Data Availability Statement}
The data that support the findings of this study are available from the corresponding author upon reasonable request.

%\bibliography{apssamp}% Produces the bibliography via BibTeX.

\end{document}